\documentclass[12pt]{article}
\usepackage{epsfig}
\usepackage{graphics}
\usepackage{cite}
\usepackage{latexsym}
\newcommand   {\uhh}    {U({\rm H},{\rm H})}  
\newcommand   {\uhp}    {U({\rm H},{\rm P})}  
\newcommand   {\upp}    {U({\rm P},{\rm P})} 

\newcommand  {\BJ}       {{\it Biophys.\ J.\ }}

\newcommand  {\JBP}      {{\it J.\ Biol.\ Phys.\ \ }}  

\newcommand  {\JCoP}     {{\it J.\ Comput.\ Phys.\ }}
\newcommand  {\JCP}      {{\it J.\ Chem.\ Phys.\ }}
\newcommand  {\JMB}      {{\it J.\ Mol.\ Biol.\ }}

\newcommand  {\Mac}      {{\it Macromolecules\ }}

\newcommand  {\Pro}      {{\it Proteins\ Struct.\ Funct.\ Genet.\ }}
\newcommand  {\ProEng}   {{\it Protein\ Eng.\ }}
\newcommand  {\ProSci}   {{\it Protein\ Sci.\ }}

\newcommand  {\PNAS}     {{\it Proc.\ Natl.\ Acad.\ Sci.\ USA\ }}
\newcommand  {\PR}       {{\it Phys.\ Rev.\ }}

\newcommand  {\SFD}      {{\it Struct.\ Fold.\ Des.\ }}

\newcommand{\beq}{\begin{equation}}
\newcommand{\eeq}{\end{equation}}
\newcommand{\beqa}{\begin{eqnarray}}
\newcommand{\eeqa}{\end{eqnarray}}
\newcommand{\bea}{\begin{eqnarray}}
\newcommand{\eea}{\end{eqnarray}}
\newcommand   {\ev}[1]   {\langle #1\rangle} 

\begin{document}

\begin{flushright}
Revised version\\
LU TP 01-07\\
November 28, 2001
\end{flushright}

\vspace{.2in}

\begin{center}

{\LARGE \bf Enumerating Designing Sequences} 

{\LARGE \bf in the HP Model}

\vspace{.3in}

\large
Anders Irb\"ack and Carl Troein\footnote{E-mail: anders,\,carl@thep.lu.se}\\ 

\vspace{0.1in}

Complex Systems Division, Department of Theoretical Physics\\
Lund University,  S\"olvegatan 14A,  S-223 62 Lund, Sweden \\
{\tt http://www.thep.lu.se/complex/}\\              

\vspace{0.25in}  

Submitted to \JBP

\end{center}
\vspace{0.25in}
\normalsize     
Abstract:\\
The hydrophobic/polar HP model on the square lattice has been widely used
to investigate basics of protein folding. In the cases where all designing
sequences (sequences with unique ground states) were enumerated without
restrictions on the number of contacts,
the upper limit on the chain length $N$ has been 18--20 because of the
rapid exponential growth of the numbers of conformations and sequences. We
show how a few optimizations push this limit by about 5 units. Based on
these calculations, we study the statistical distribution of hydrophobicity
along designing sequences. We find that the average number of hydrophobic
and polar clumps along the chains is larger for designing sequences than
for random ones, which is in agreement with earlier findings for $N\le18$
and with results for real enzymes. We also show that this deviation from
randomness disappears if the calculations are restricted to maximally compact
structures.
 
\vspace{12pt}
 

\newpage

\section{Introduction}

Coarse-grained models are an important tool in theoretical studies 
of protein folding, for computational as well as conceptual reasons, 
and have been used to gain insights into the physical principles 
of folding (for a recent review, see Ref.~\cite{Chan:00}). These models 
are often lattice based. 
The main advantage of using a discrete conformational space 
is that exact calculations can be performed for short chains,
by exhaustive enumeration of all possible conformations. One model
that has been extensively studied this way is the minimalistic 
hydrophobic/polar HP model of Lau and Dill~\cite{Lau:89} on the 
square lattice. In previous work on this model, all sequences
with unique ground states were determined for chains 
with up to $N=18$--20 monomers~\cite{Dill:95,Hirst:99,Shahrezaei:00}. 
Such sequences are called designing; they design their
ground state conformations. 

In this paper, we show how a few optimizations make it 
possible to extend these calculations to $N=25$, which corresponds 
to a 4000-fold increase in the number of possible sequence, conformation 
pairs.\footnote{The complete list of all designing $N\leq25$ sequences and 
the corresponding conformations will be made electronically 
available at {\tt http://www.thep.lu.se/complex/wwwserver.html}}   
We then use this data set to address the question of how 
designing sequences differ from random ones statistically. 

By analyzing the behavior of block variables, it has been found that 
designing $N\le18$ HP sequences~\cite{Irback:00} as well as
real (globular) protein sequences~\cite{Irback:00,Irback:96}  
show negative hydrophobicity correlations. Therefore, 
one expects to find an increased number of hydrophobic 
and polar clumps along these chains. In this paper,
we show that the average number of clumps is indeed larger for 
designing HP sequences than for random sequences. In particular, 
this implies that the earlier finding that designing sequences
show negative hydrophobicity correlations remains unaffected 
when increasing $N$ to 25. This provides a non-trivial test of   
the robustness of this property. 

In lattice model studies it is not uncommon to consider only
maximally compact conformations, which for $N=25$ are
confined to a $5\times5$ square. This drastic reduction 
of conformational degrees of freedom leads to a sharp rise
in the number of designing sequences. An interesting question
is whether this reduction also affects the statistical 
properties of designing sequences. To address this question,
we repeat the same statistical analysis for  
sequences that are designing when only maximally compact 
conformations are used. The results turn out to be qualitatively different 
in this case. In particular, this means that the agreement with the results 
obtained for real sequences gets lost when this reduced  
conformational space is used. 

Finally, we also study the character of the folding transition
for one of the designing $N=25$ sequences, which was selected 
by an optimization procedure. The thermodynamic behavior of this
sequence is studied using Monte Carlo simulations. 

The paper is organized as follows. In Section~\ref{sec:algo} we define the 
model and describe the algorithm and optimizations used for finding all 
designing sequences with $N\le 25$. Our results are discussed 
in Section~\ref{sec:res}, which contains sequence and structure statistics, 
the statistical analysis of designing sequences, and the  
thermodynamic study of an optimized sequence. A summary is given in 
Section~\ref{sec:sum}.     

\section{Enumerating designing sequences}
\label{sec:algo}

In lattice models of proteins it is common to use a contact potential. This
means that the energy that a sequence gets with a certain conformation, is
given by what contacts exist in that conformation. That is,
\beq 
E = \sum_{i<j} C_{ij}U(\sigma_i,\sigma_j)
\label{E}\eeq
where the contact map $C_{ij}$ is defined as
\beq 
C_{ij} = \left\{ \begin{array}{rl}
1 & \textnormal{if monomers $i,j$ are neighbors on the lattice but\ }
|i-j| \neq 1\\ 0 & \textnormal{otherwise}
\end{array} \right. 
\label{Cij}\eeq
and $U(\sigma_i,\sigma_j)$ is the interaction matrix. In the HP model there
are two amino acids, hydrophobic (H) and polar (P). The interaction matrix is
\beq 
U = \left( \begin{array}{rr}-1 & 0 \\ 0 & 0\end{array} \right) 
\label{U}\eeq
with H in the first row and column, so the energy is determined exclusively
by the number of HH contacts. 

Two conformations that have identical contact sets will 
have the same energy for all sequences, so they can be
represented by a single contact set which is marked as {\it degenerate}.
What this means is that a large number of conformations can be reduced to a
smaller number of contact sets. A contact set that corresponds to a single
conformation will be called {\it unique}.

\subsection{Compressing conformational space}

To find all designing sequences, we first determine all relevant 
conformations, which then are combined with sequences
in Section~\ref{sec:seq}. The conformation search is sometimes
simplified by only considering maximally compact conformations, where
the number of contacts is maximal. Looking only at those conformations
corresponds to shifting the energies by adding a large negative term to all
elements of the interaction matrix. An efficient method for enumerating
compact conformations was recently proposed by Kloczkowski and
Jernigan~\cite{Kloczkowski:98}.   

In this paper, we consider all possible self-avoiding walks, 
and not only those that are maximally compact. 
The space of all possible self-avoiding walks grows rapidly with $N$.
Using contact sets rather than self-avoiding walks gives a significant 
reduction of the conformational space (see Table~2 below), but this reduction 
is not sufficient for our purposes; memory limitations 
prevented us from storing the complete list
of all possible contact sets for $N\ge24$. To be able to go to larger 
$N$, we therefore developed two procedures for reducing the number of 
contact sets to be stored. These procedures are based on the observation 
that as long as there are no repulsive forces (that is, as long as the 
elements of the interaction matrix are all non-positive), it is never 
energetically disadvantageous to add contacts as long as no existing 
contacts are broken. 

Before discussing these two procedures, it is helpful to introduce 
some terminology. A contact set will be called  {\it maximal} if it 
is not a proper subset of any other (realizable) contact set. A
conformation that is designed by at least one sequence is designable.
It can be readily seen that every designable 
self-avoiding walk corresponds to a unique and maximal
contact set. Another important class are contact sets    
that can be safely ignored in the search for designing sequences. 
Such contact sets will be called {\it redundant}. A summary of our 
contact set terminology can be found in Table~1.        

\begin{table}[t]
\caption{Contact set terminology.}
\begin{center}
\begin{tabular}{ll}
\hline
Unique & Corresponds to a single self-avoiding walk \\ 
Degenerate & Corresponds to more than one self-avoiding walk \\
Maximal & Not a subset of any other realizable contact set\\
Redundant & Can be ignored in the search for designing sequences\\
\hline
\end{tabular}
\end{center}
\end{table}

\subsubsection{Eliminating contact sets: Step 1} 
\label{sec:list}

As mentioned above, it was impossible for us to store the complete list
of all contact sets for $N\ge24$. To circumvent this problem, 
we used a program that for each self-avoiding walk tries a 
carefully selected, predefined set of (local) moves. If any of these 
moves can be 
performed without destroying any existing contact (new contacts may
form), the self-avoiding walk is discarded. All possible self-avoiding 
walks surviving this test are converted to contact sets. It is important
to stress that the moves are chosen so that 
the resulting reduced list of contact sets has 
the property that any realizable contact set is a subset of some set 
in this list. In particular, the move set does not 
contain the inverse of any of its elements.

The list of contact sets obtained this way was indeed small enough 
to be stored up to $N=25$, but the procedure has the disadvantage 
that it may eliminate non-redundant contact sets. This is an unwanted
property, but the problem is easy to solve.  
The solution is that once the sequences 
that have non-degenerate ground states 
{\em with respect to the non-discarded conformations} 
have been found, each sequence is combined with its
conformation, and the chain is tested with the opposite of the tests used
to discard conformations. That is, the program tests whether 
any of the opposite moves can be performed without breaking 
any of the existing HH contacts. By using the fact that the forces
are repulsive, it can be seen that if no such move is possible,   
then the conformation has to be a unique ground state of 
this sequence. Hence, by performing this test, one can make sure 
that no sequence is falsely declared to be designing, even though 
there are non-redundant contact sets that are missing in the list used.    

It is also important to note that all the self-avoiding walks with a 
given contact set are never discarded if the set is maximal. This
means that all the contact sets that correspond to designable
conformations are included in the list generated by this procedure.
This is important because it implies that this reduced list of
contact sets can be used without losing any designing sequence.
What could go wrong when using this list is that some non-designing
sequences were classified as designing, but this is avoided by 
performing the test discussed above.   

\subsubsection{Eliminating contact sets: Step 2}\label{sec:elim}

Our second procedure removes contact sets from the list produced by 
the first procedure. This is done to speed up the calculations. 
All contact sets that are removed in this second step are redundant.   

The procedure relies on the fact that all pair energies are non-positive.
To see how that can be used, consider one set of contacts $A$ 
which is a subset of another contact set $B$. It is then clear that $A$ 
cannot represent a unique ground state, because for any sequence, 
$B$ has the same or lower energy. The set $A$ is nevertheless non-redundant 
in case $B$ is a unique contact set that would be falsely classified
as the unique ground state of some sequence if $A$ were removed.
If, on the other hand, the set $B$ is degenerate, then it follows that    
$A$ must be redundant. 

Suppose now that $A$ is a subset of two other sets $B$ and $C$. Then,  
provided that $C$ is kept, there can be no sequence such that $A$ is 
needed in order to decide whether or not this sequence has $B$ as a unique 
ground state. This follows immediately from the fact that for any given 
sequence, both $B$ and $C$ have energies at least as low as that of $A$. 
Hence, if a contact set is a subset of two or more sets, then it has  
to be redundant. In particular, this is true if the set 
is a subset of a subset.

This reasoning gives us the following simple procedure for elimination  
of redundant contact sets. 
\begin{itemize}
\item For each set $A$, find all sets of which $A$ is a subset.
\item Keep $A$ if
\begin{enumerate}
\item no such sets are found, or
\item {\em one} set is found, and this set corresponds to {\em one}
  conformation.
\end{enumerate}
\item Otherwise discard $A$.
\end{itemize}
Note that those of the surviving contact sets that meet condition 1 
are maximal.

It should be stressed that, because all the contact sets removed
by this procedure are redundant, one can still use the test in 
Section~\ref{sec:list} to make sure that no sequence is 
falsely classified as designing.    

Implementing the rules described above requires some care, since the
number of sets grows rapidly with $N$. The solution we used is based 
on storing the sets in a tree, where each node in the tree represents 
the question of whether or not a certain contact is included.
Before building the tree, it is useful to sort the sets in 
descending order by the number of contacts they contain. This sorting 
ensures that a set, once added to the tree, never has to be removed, since
all sets of which it can be a subset are already added when it is examined.
Having access to the tree, it is straightforward to search 
for supersets of a given set. One starts at the root of the
tree and each time the tree branches one has to consider either one of the
branches or both. The procedure is memory-consuming for large $N$ (see example 
below), but the CPU time required is relatively modest.

\subsubsection{Examples}\label{sec:ex}

To get an idea of the sizes of these different lists of contact sets,
let us consider the case $N=18$. 
For this $N$, there are 5808335 self-avoiding walks and
170670 contact sets. After applying the redundancy test in 
Section~\ref{sec:elim} to the list of all possible contact sets,  
we are left with 33223 maximal contact sets  
(condition 1) and 6609 contacts sets with one superset 
corresponding to one conformation (condition 2). Among the
33223 maximal contact sets, there are 6181 sets representing 
more than one conformation. Subtracting the degenerate contact sets, 
we are left with $33223 - 6181 = 27042$ contact sets corresponding to 
possibly designable conformations. Of those 27042 sets, 5660 sets have 
a total of 6609 listed subsets. These subsets are needed because they 
may degenerate an otherwise unique ground 
state corresponding to one of the 5660 sets.

Here, the redundancy test was applied to the complete list of all 
possible contact sets. Alternatively, we may first use the program 
described in Section~\ref{sec:list}, which for $N=18$ generates 
a list of 51373 contact sets (which contains the 33223 maximal ones). 
When applying the redundancy test in Section~\ref{sec:elim} 
to this reduced list, we end up with $33223+449$ contact sets. 
Note that with this approach, 
we find only 449 of the 6609 non-redundant 
contact sets meeting condition 2. 
As a result, some conformations are erroneously 
found to be unique ground states. The test discussed in 
Section~\ref{sec:list} removes these false unique ground states. 

The CPU times required to generate these different lists on a 
Pentium III 800\,MHz were as follows. Generating the list 
of all possible contact sets, by 
exhaustive enumeration, took 6 seconds, and reducing this list
from 170670 to $33223+6609$ contact sets required 7 seconds. 
The time needed to run the program that generates the list described 
in Section~\ref{sec:list} was 3 seconds, and reducing 
this list from 51373 to $33223+449$ contact sets took 1 second.  
The corresponding two numbers for $N=25$ were 40 and 60 minutes,
respectively. In this case, building the tree used in the  
redundancy test required 220 megabytes of internal memory. 

\subsection{Searching sequence space}\label{sec:seq}

We now turn our attention to the sequence space. 
For each of the $2^{N}$ sequences we wish to determine what set of contacts
gives the lowest energy. If this ground state energy can be accomplished
only by a single contact set, and if that set corresponds to a single
conformation, the sequence designs that conformation.

\subsubsection{Organizing the search}

The most straightforward approach to finding all sequences with unique
ground states is to go through all the sets of contacts for each sequence,
and calculate the energy for each of the sets. 
By only storing the differences
between consecutive sets of contacts, and by representing the sequences and
contacts as numbers with one bit per position, the number of operations
required for each combination can be kept small.

It is, however, also possible to use a very different approach. Represent
each sequence by a binary number, and consider any given set of contacts.
Between two consecutive sequences two bits are toggled on the average,
which indicates that using information about the previous sequence and its
energy will be a lot faster than recalculating the energy from scratch.
This approach can be used if the whole sequence space is examined for one
contact set at a time. The downside to doing this is that information needs
to be stored for every sequence until all the contact sets have been
considered.

Neither of the methods described above is bad, but they each contain an
optimization that the other does not have. It is desirable to utilize both
the similarity of consecutive sequences and the similarity of consecutive
contact sets. The solution is to divide the sequence space into a number of
blocks of fixed size, and apply the second method to each of those blocks.
A block consists of $2^{M}$ sequences that have their first $N-M$ residues
in common. This part of the sequence will be referred to as the {\em fixed
part}, and the remaining $M$ positions make up the {\em variable part}. We
call a contact {\em active} if it connects two H monomers, and note that
for each contact there are three possibilities, depending on the position
and type of the monomers it connects:
\begin{itemize}
\item If both monomers are in the fixed part, the contact gives a contribution
of $-1$ to the energy for {\em all} the sequences of this block, if and only
if both monomers are H.
\item If at least one of the monomers is a P in the fixed part, the contact
cannot be active for any sequence in the block.
\item If both monomers are in the variable part, or if one of them
is in the variable part and the other is an H in the fixed part, whether or
not the contact is active depends on the variable part.
\end{itemize}

\subsubsection{The cutoff energy}
\label{sec:cutoff}

This leads us to the next optimization, which has to do with the possible
ground state energies, and can be seen as a sequence-dependent reduction of
the conformational space. Clearly, a non-degenerate ground state with $N
\geq 3$ cannot have an energy of~0. More generally, it is unreasonable that
a small number of active contacts should be enough to give an arbitrarily
long polymer a unique ground state.
To see how this can be used to speed up the calculations, consider some
contact set and sequence block, such that there are $p$ active contacts in
the fixed part and $q$ contacts whose state depends on the variable part.
If $-(p+q)$ is larger than some cutoff value $E_{\max}$, none of the sequences
in this block can have a unique ground state for this contact set. The
major problem that arises with this optimization is to know what value to
use for $E_{\max}$; the algorithm will find only those unique ground states
that have energies $E\le E_{\max}$. 
For $N\leq20$ all energies have been considered in the calculations, and
it turns out that there are no unique ground states with $E>-4$ 
for $15\leq N\leq 20$ (see Table~3 below). For $N>20$ we have not
proven that there can be no unique ground states with $E > -4$, but it
seems very reasonable that {\em if} there are any, most or all of them
would have $E = -3$. Therefore we have used a cutoff energy $E_{\max} = -3$
for $N > 20$. It turns out that for $20<N\leq25$ there are no unique ground
states with energy $-3$, and this strongly indicates that $-4$ is the
highest possible ground state energy for any HP chain with $N \geq 15$.

To illustrate how the computer time required varies with $E_{\max}$,
let us again take $N=18$ as an example. For this $N$, the sequence search  
is found to be about four times faster with $E_{\max} = -4$ than 
with $E_{\max} = -1$. With $E_{\max}=-4$, the total time needed to 
find the 6349 designing $N=18$ sequences, including the time required 
to generate the conformations, is a few minutes on a Pentium III 800\,MHz. 

\newpage

\section{Results}
\label{sec:res}

Using the algorithm and optimizations described above, it was possible to
determine all designing sequences for $N \leq 25$ within a reasonable amount 
of time. Previous work has covered $N\leq18$ for the normal HP 
model~\cite{Dill:95} and $N\leq20$ for shifted HP 
models~\cite{Hirst:99,Shahrezaei:00}. The increase in $N$ 
corresponds roughly to a 100-fold increase in the number of
known designing sequences and conformations. This gives better confidence
when doing statistics on the designing sequences, and it makes it possible
to study how properties of the model depend on protein size.

\subsection{Sequence and structure statistics}

Sequence and structure statistics for $4 \leq N \leq 25$ are summarized in 
Tables~2 and 3. 
Column three of Table~2 shows the total number of contact sets, which
has been studied before~\cite{Vendruscolo:99}. It was estimated to 
grow as $\mu^N$ with $\mu=2.29\pm0.02$ for large $N$~\cite{Vendruscolo:99}. 
The number of maximal contact sets, column four of Table~2, 
appears to grow exponentially too, but
slightly slower. A fit of our data for $15\le N\le25$ to the
form $\mu^N$ yields $\mu\approx 2.07$. This growth with $N$ is considerably
slower than that of the number of self-avoiding walks, for which the  
best available estimate is $\sim N^{\gamma-1}\mu^N$
with $\gamma=43/32$ and $\mu=2.6381585$ \cite{Madras:93}.   

The fifth column of Table~2 shows the number of designing sequences.
It turns out that the fraction of designing sequences  
varies between 2.27 and 2.57\% for $19\leq N\leq25$, 
which is in line with previous results for smaller 
$N$~\cite{Chan:94}. The last column of Table~2 shows the number
of designable conformations. The designability of a conformation
is the number of sequences that designs it. From Table~2 it can be
seen that the average designability of the conformations that are 
designable grows with $N$ and is $765147/107336\approx 7.1$ for $N=25$.

In Table~3 we show the distribution of ground state energies for 
different $N$, which is crucial for the optimization 
discussed in Section~\ref{sec:cutoff}.     

\begin{table}[t]
\small
 \caption{\small 
 The number of designing sequences $S_N$ and designable
 conformations $D_N$ for the HP model on a square lattice. The third
 column shows the total number of contact sets, as obtained by exhaustive
 enumeration of all possible self-avoiding
 walks. Memory requirements (see Sec~\ref{sec:list}) prevented us from counting
 them for $N\geq 24$. Column four shows the number of maximal contact
 sets.
 }
\begin{center}
 \begin{tabular}{r@{\extracolsep{24pt}}r@{}r@{}r@{}r@{}r}
 \hline\hline
 &&& Maximal \\
 $N$ & Conformations & Contact sets & contact sets & $S_N$ & $D_N$ \\
 \hline
 4 &          5 &        2 &       1 &      4 &      1 \\
 5 &         13 &        3 &       2 &      0 &      0 \\
 6 &         36 &        8 &       4 &      7 &      3 \\
 7 &         98 &       14 &       9 &     10 &      2 \\
 8 &        272 &       41 &      20 &      7 &      5 \\
 9 &        740 &       78 &      39 &      6 &      4 \\
10 &       2034 &      212 &      95 &      6 &      4 \\
11 &       5513 &      424 &     174 &     62 &     14 \\
12 &      15037 &     1113 &     420 &     87 &     25 \\
13 &      40617 &     2309 &     779 &    173 &     52 \\
14 &     110188 &     5953 &    1818 &    386 &    130 \\
15 &     296806 &    12495 &    3409 &    857 &    218 \\
16 &     802075 &    31940 &    7810 &   1539 &    456 \\
17 &    2155667 &    67389 &   14717 &   3404 &    787 \\
18 &    5808335 &   170670 &   33223 &   6349 &   1475 \\
19 &   15582342 &   363010 &   63434 &  13454 &   2726 \\
20 &   41889578 &   910972 &  140939 &  24900 &   5310 \\
21 &  112212146 &  1953847 &  273049 &  52183 &   9156 \\
22 &  301100754 &  4868343 &  599821 &  97478 &  17881 \\
23 &  805570061 & 10513774 & 1174460 & 199290 &  31466 \\
24 & 2158326727 &          & 2561884 & 380382 &  61086 \\
25 & 5768299665 &          & 5057733 & 765147 & 107336 \\  
 \hline\hline
 \end{tabular}
\end{center}
\end{table}

\begin{table}[htb]
\scriptsize
 \caption{\small The number of designing sequences for different $N$ and
 ground state energies.
 }
\vspace{6pt}
 \begin{tabular}{r@{\extracolsep{8.5pt}}r@{}r@{}r@{}r@{}r@{}r@{}
  r@{}r@{}r@{}r@{}r@{}r@{}r@{}r@{}r@{}r}
 \hline\hline
 $N$ & \multicolumn{16}{l}{Energy} \\
  & -1 & -2 & -3 & -4 & -5 & -6 & -7 & -8 & -9 & -10 & -11
   & -12 & -13 & -14 & -15 & -16 \\
 \hline
4 & 4 \\
 5 \\
 6 & 0 & 7 \\
 7 & 0 & 10 \\
 8 & 0 & 0 & 7 \\
 9 & 0 & 0 & 8 \\
 10 & 0 & 0 & 1 & 5 \\
 11 & 0 & 0 & 6 & 54 & 2 \\
 12 & 0 & 0 & 2 & 27 & 49 & 9 \\
 13 & 0 & 0 & 0 & 78 & 54 & 41 \\
 14 & 0 & 0 & 2 & 53 & 110 & 132 & 89 \\
 15 & 0 & 0 & 0 & 58 & 88 & 355 & 330 & 26 \\
 16 & 0 & 0 & 0 & 43 & 158 & 250 & 638 & 417 & 33 \\
 17 & 0 & 0 & 0 & 33 & 160 & 662 & 882 & 1337 & 330 \\
 18 & 0 & 0 & 0 & 24 & 149 & 623 & 1431 & 2021 & 1676 & 425 \\
 19 & 0 & 0 & 0 & 8 & 154 & 971 & 1936 & 4996 & 3324 & 2007 & 58 \\
 20 & 0 & 0 & 0 & 13 & 147 & 955 & 2573 & 5582 & 7665 & 5415 & 2481 & 69 \\
 21 &&& 0 & 17 & 134 & 1312 & 3116 & 11670 & 12132 & 13917 & 8898 & 987 \\
 22 &&& 0 & 12 & 120 & 1116 & 3802 & 11672 & 22386 & 24171 & 22394 & 10610 &
  1195 \\
 23 &&& 0 & 26 & 92 & 1547 & 4204 & 21050 & 29944 & 56902 & 44940 & 31961 &
  8118 & 506 \\
 24 &&& 0 & 17 & 134 & 1321 & 4916 & 21096 & 48017 & 78496 & 100746 & 75346 &
  40376 & 9596 & 321 \\
 25 &&& 0 & 20 & 64 & 1708 & 5270 & 32484 & 59470 & 158044 & 159704 & 191377 &
  102944 & 46386 & 7688 & 6 \\                                                 
 \hline\hline
 \end{tabular}
\end{table}

Figure~1 shows three designable $N=25$ conformations. The conformation
(a) is the most designable one for $N=25$, with a designability of 326, 
whereas conformation (c) is designed by one sequence only.

\begin{figure}[t]
\hspace{3mm}
\resizebox{14.0cm}{!}{
\begin{picture}(130,130)
\put(0,0){\epsfig{figure=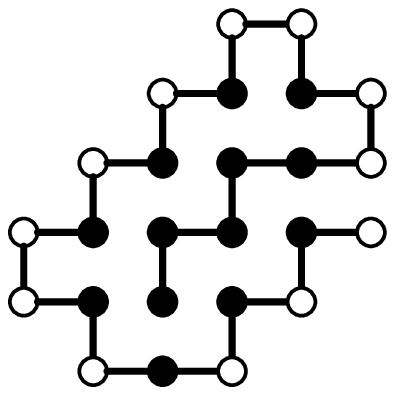}}
\put(5,105){(a)}
\end{picture}
\hspace{4mm}
\begin{picture}(110,130)
\put(0,0){\epsfig{figure=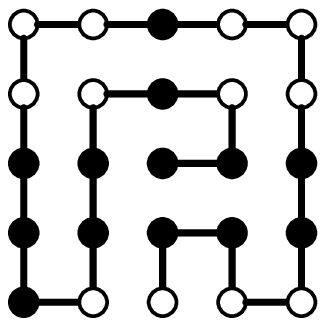}}
\put(5,105){(b)}
\end{picture}
\hspace{4mm}
\begin{picture}(170,130)
\put(0,0){\rotatebox{90}{\epsfig{figure=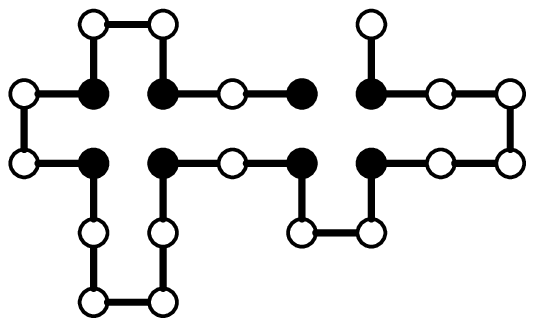}}}
\put(5,105){(c)}
\end{picture}
}
\caption{\small 
Some designable structures with $N=25$: (a) the most
designable structure, (b) a maximally compact structure, and (c) a
structure with few contacts. Filled circles represent H.
}
\end{figure}

For comparison, we also determined the sequences that are designing  
when only maximally compact conformations are used. In this case, 
it turns out that there are 6181800 designing $N=25$ sequences. The 
corresponding number is 765147 when the full conformational 
space is used (see Table~2), so the ratio between the number 
of designing sequences in the maximally compact ensemble
and the number of truly designing sequences is 
$6181800/765147\approx 8.1$, for $N=25$. This ratio has previously 
been shown~\cite{Chan:96} to fluctuate between
approximately 4 and 11 for $N=11$ through $N=18$.

It is worth noting that among the 765147 $N=25$ sequences that are
designing when the full conformational space is used, there are only
605 sequences that design maximally compact conformations. Furthermore,
it turns out that no maximally compact conformation is designed by 
more than 10 sequences, whereas the most designable conformation, as  
mentioned above, has a designability of 326. In fact, there are 19360 
conformations that are more designable than the most designable one
among the maximally compact conformations.

It is interesting to compare these results to those of 
Shahrezaei and Ejtehadi~\cite{Shahrezaei:00}, who studied a
shifted HP model where the interaction matrix is given
by $\uhh=-2-\gamma-E_c$, $\uhp=-1-E_c$ and $\upp=-E_c$.
Using the full conformational ensemble, these authors 
found that the set of highly designable conformations 
was independent of the parameters $\gamma$ and $E_c$.   
In particular, this suggests that the set of highly designable 
conformations should remain the same when
only maximally compact conformations are considered. Our results
show that this conclusion does not hold in the original, unshifted
HP model.       

\subsection{Statistical properties of designing sequences}
\label{hydro}

In this section, we study the statistical properties of designing 
sequences by monitoring two different quantities. The first one is 
the total hydrophobicity $M$, defined as
\beq
M=\sum_{i=1}^N\frac{1+\sigma_i}{2}\,,
\eeq
where $\sigma_i=1$ (H) or $\sigma_i=-1$ (P). Our second quantity 
is the number $j$ of hydrophobic and polar clumps along the 
chain~\cite{White:90}, which can be written as
\beq  
j = \frac{N+1}{2} - \frac{1}{2} \sum_{i=1}^{N-1} \sigma_i\sigma_{i+1}\,. 
\eeq
A similar analysis has previously been performed for 
$N\le 18$~\cite{Irback:00}. By analyzing the fluctuations 
of block variables, evidence was found for
negative $\sigma_i,\sigma_j$ correlations. Therefore, we
expect the average $j$ to be larger for designing sequences 
than for random ones. 

In Figure~2a we show the relative abundance of 
hydrophobic amino acids, $\ev{M}/N$, as a function of $N$, where
$\ev{\cdot}$ denotes an average for fixed $N$.
For the sequences
that are designing when the full conformational space is used, 
we see that the $N$ dependence of $\ev{M}/N$ is 
quite weak if $N$ is not too small, which confirms a trend seen 
earlier~\cite{Irback:00}. Furthermore, we see that these sequences,
as expected, are more hydrophobic than those that are designing when 
only maximally compact conformations are used. Finally, it can also
be seen that the sequences that design maximally compact conformations 
differ greatly from those that are 
designing when only such conformations are used. 
Figure~2b shows the frequency of different $M$ for $N=25$.       

\begin{figure}[t]
\hspace{6mm}
\resizebox{7.1cm}{!}{\rotatebox{270}{\epsfig{figure=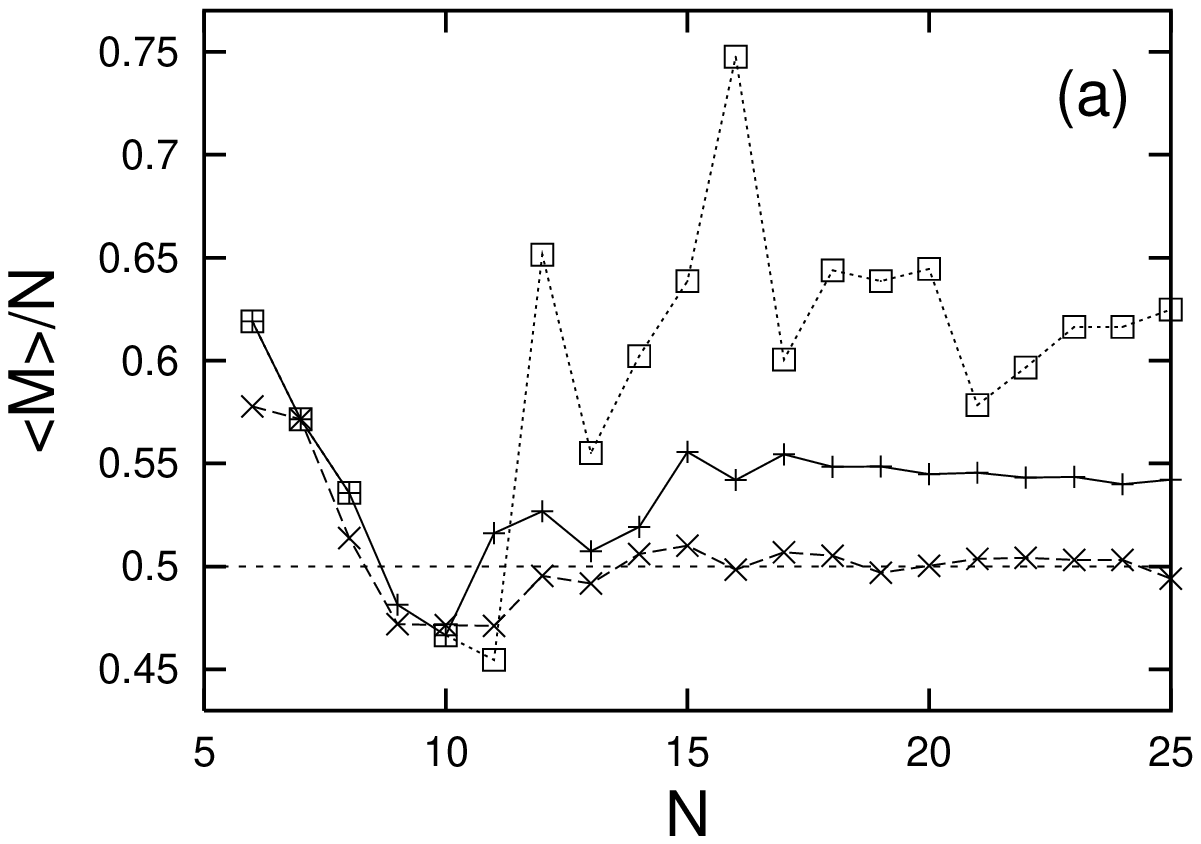}}}
\resizebox{7.1cm}{!}{\rotatebox{270}{\epsfig{figure=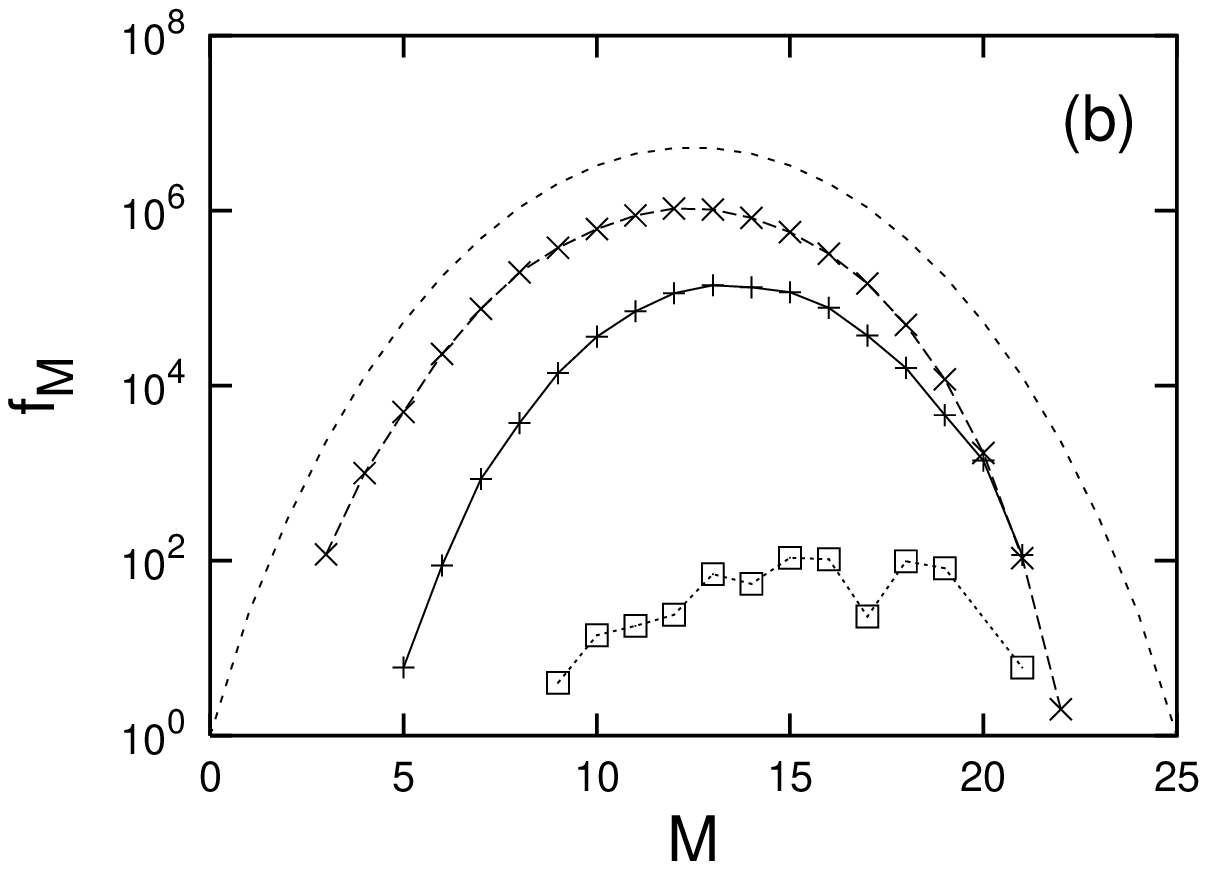}}}
\caption{\small 
 (a) The average hydrophobicity $\ev{M}/N$ as a function
 of $N$ and (b) the frequency $f_M$ of different $M$ for $N=25$. Shown are 
 the results for all designing sequences ($+$), all sequences that
 are designing when only maximally compact conformations are considered
 ($\times$), and sequences
 that design maximally compact conformations when all conformations are
 considered ($\Box$). The dashed lines represent random sequences.
}
\end{figure}

In Figure~3 we show the results of our clump analysis 
for $N=25$. The average number of clumps for fixed $M$ 
(and $N$), $\ev{j}_M$, is indeed found to be larger 
for designing sequences than for random ones, which is in 
nice qualitative agreement with previous results for real
protein sequences and model sequences with smaller 
$N$~\cite{Irback:00,Irback:96}. Sequences that are designing when the 
interaction energies are shifted so far that 
only the 1081 maximally compact conformations need
to be considered, have, by contrast, a $\ev{j}_M$ close to that of 
random sequences. Hence, the results obtained from studying 
only maximally compact conformations seem to be of less relevance, 
with respect to the applicability to real proteins, than the
results obtained when considering all conformations.

\begin{figure}[t]
\hspace{33mm}
\resizebox{7.1cm}{!}{\rotatebox{270}{\epsfig{figure=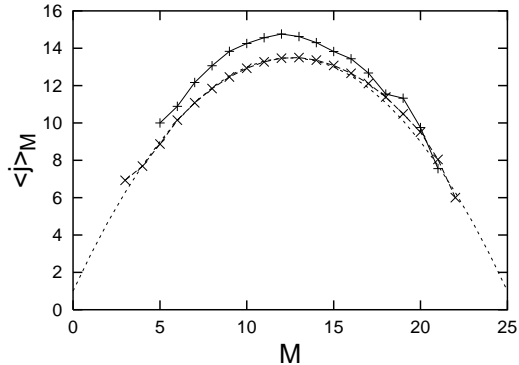}}}
\caption{\small 
 The number of clumps, $\ev{j}_M$, as a function of the 
 total hydrophobicity $M$ for $N=25$. Shown are the results for sequences 
 that are designing when all conformations ($+$) and only maximally compact 
 ones ($\times$), respectively, 
 are considered. The dashed line represents 
 random sequences.
}
\end{figure}

Finally, we note that Buchler and Goldstein have compared 
various lattice models and found arguments against the use of only 
two letters~\cite{Buchler:99,Buchler:00}, as in the HP model. 
These findings were all based on calculations for maximally compact 
conformations. However, it is not known to what extent the results of 
such calculations remain valid when the full set of conformations
is used. Our results show that, in the HP model, both the set of 
highly designable conformations and the statistical properties 
of designing sequences depend strongly on which of the two 
conformational ensembles is used.  
 
\subsection{The character of the folding transition}

The model studied in this paper has the important feature that there 
exists a significant number of designing sequences (this is not true  
on the triangular~\cite{Irback:98} and cubic~\cite{Yue:95} lattices), 
and that the corresponding conformations tend to show protein-like 
regularities~\cite{Chan:90}. However, as in other two-dimensional
models, the folding transition is not protein-like in character
for the typical sequence; the folding process is not 
cooperative enough~\cite{Abkevich:95}. On the other hand, 
the folding behavior is, at least to some extent, sequence dependent,
and therefore we decided to look into the thermodynamic behavior      
of a carefully chosen sequence. 

This sequence was obtained by applying a Monte Carlo-based 
sequence design algorithm~\cite{Irback:99} to the 
326 sequences that design the most designable 
$N=25$ conformation (see Figure 1a). The design algorithm maximizes 
the stability of a given conformation with respect to sequence 
at a fixed non-zero temperature. The sequence we obtained by
using this method is shown in Figure~1a. Subsequently, 
this sequence was subjected to Monte Carlo simulations at 
different temperatures. 
In Figure~4 we show the temperature-dependence of the specific heat, 
which is found to exhibit a pronounced peak. Also shown is  
the distribution of energy at $kT=0.479$, which is
just above the specific-heat maximum. 
The energy distribution has one peak corresponding
to the ground state, at $E=-13$, and another, broader peak centered at
$E\approx -8$. The coexistence of these states implies that the
folding transition is much more cooperative for this sequence than
for the typical sequence in this model.

\begin{figure}[t]
\hspace{0mm}
\resizebox{7.1cm}{!}{\rotatebox{270}{\epsfig{figure=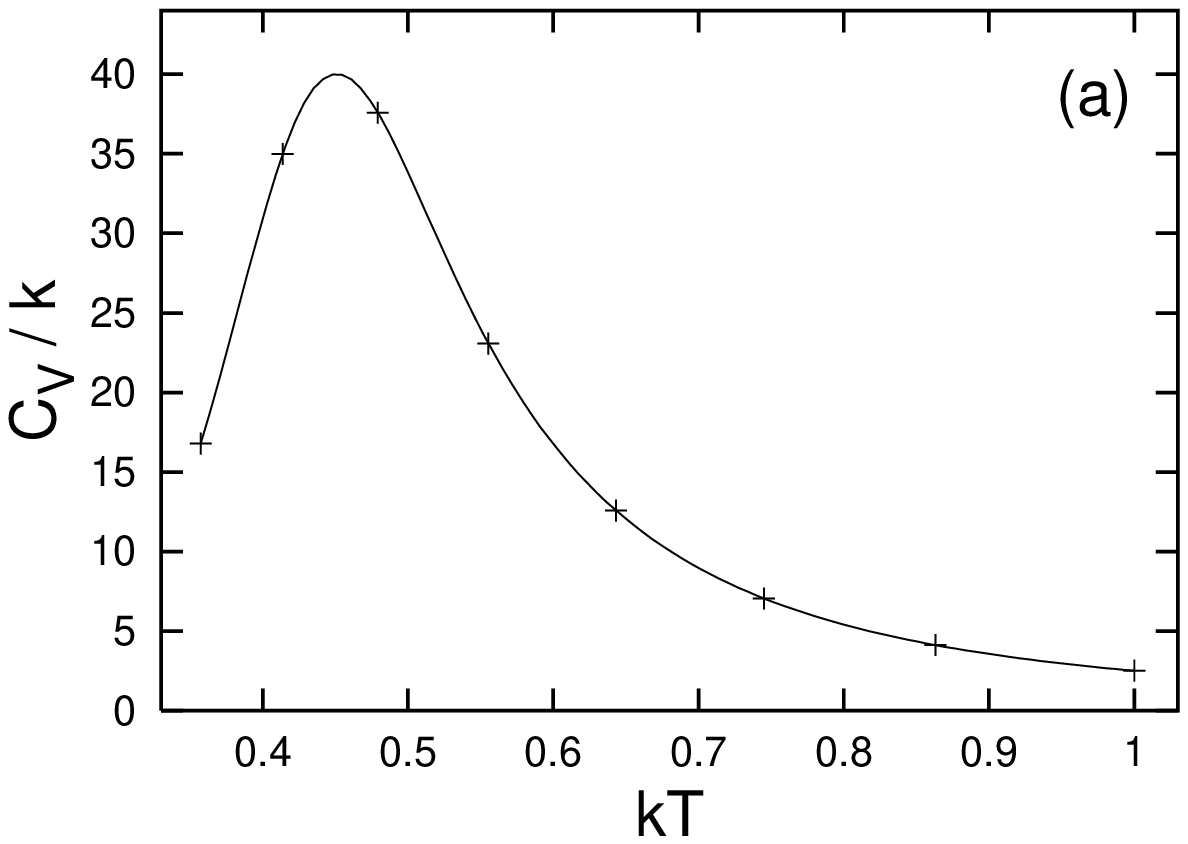}}}
\resizebox{7.1cm}{!}{\rotatebox{270}{\epsfig{figure=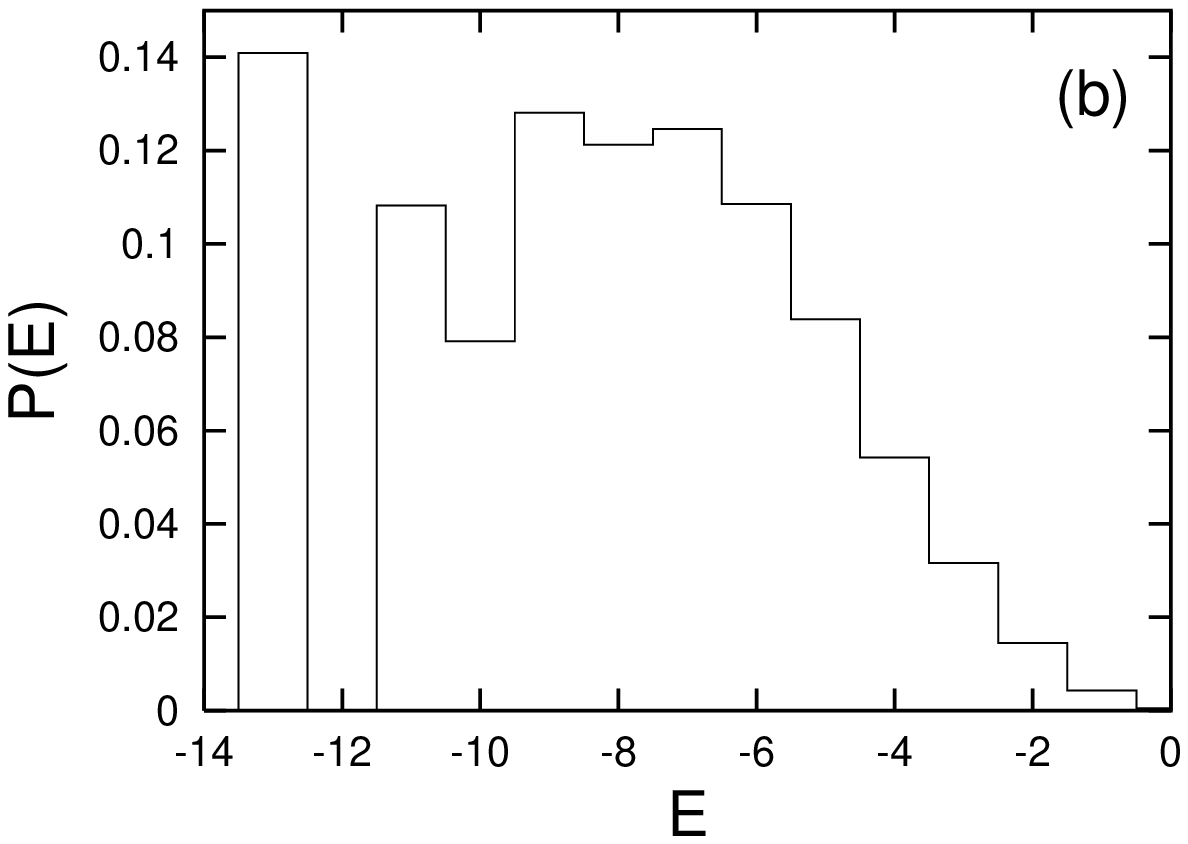}}}
\caption{\small 
Results from Monte Carlo simulations of the 
sequence shown in Fig.~1a. (a) Temperature
dependence of the specific heat $C_v=(\ev{E^2}-\ev{E}^2)/kT^2$. 
The line is an extrapolation obtained by umbrella 
sampling~\cite{Torrie:77}. 
(b) Histogram of energy at $kT=0.479$.
}
\end{figure}

\section{Summary}
\label{sec:sum}

By greatly reducing the conformational space, and by carefully optimizing
the sequence space exploration, we were able to decrease the time needed to
exhaustively search for designing sequences with $N=18$ roughly
thousandfold compared to the most na\"{\i}ve methods. It became feasible to
find all designing sequences for $N$ as large as 25 using only a small 
number of workstations. The results obtained by doing this were used to 
look at the statistical properties of designing sequences. We found that the
average number of hydrophobic and polar clumps along the chains is larger
for designing sequences than for random ones. In particular, this means that  
the finding that designing HP sequences, like real enzymes, 
show negative hydrophobicity correlations~\cite{Irback:00,Irback:96}
remains unaffected when increasing $N$ from 18 to 25.
By contrast, qualitatively different results were obtained
when discarding conformations that are not maximally compact. This is of 
interest because restrictions to compact structures are common
in both lattice model studies and determinations of statistical 
potentials from known protein structures. Finally, we saw an example 
of a folding behavior that is more cooperative than for the typical 
sequence in this model.

\subsection*{Acknowledgements}

We would like to thank Erik Sandelin for fruitful discussions,  
Bj\"orn Samuelsson for generously  
providing the program discussed in Section~\ref{sec:list},
and two anonymous referees for useful remarks. 
This work was in part supported by the Swedish Foundation for Strategic
Research.

\end{document}